\documentclass[letterpaper]{article} 
\usepackage{aaai2026}  
\usepackage{times}  
\usepackage{helvet}  
\usepackage{courier}  
\usepackage[hyphens]{url}  
\usepackage{graphicx}  
\usepackage{natbib}  
\usepackage{caption}  
\usepackage{amsmath}
\usepackage{amssymb}
\usepackage{enumitem}
\usepackage{booktabs}
\usepackage{multirow}
\usepackage{tcolorbox}

\usepackage{algorithm}
\usepackage{algorithmic}
\usepackage{newfloat}
\usepackage{listings}
\floatstyle{ruled}
\newfloat{listing}{tb}{lst}
\floatname{listing}{Listing}

\newcommand{\eg}{\emph{e.g., }}

\title{Navigating Through Paper Flood: Advancing LLM-based Paper Evaluation through Domain-aware
Retrieval and Latent Reasoning}
\author {
    Wuqiang Zheng\textsuperscript{\rm 1},
    Yiyan Xu\textsuperscript{\rm 1}, 
    Xinyu Lin\textsuperscript{\rm 2}, 
    Chongming Gao\textsuperscript{\rm 1},
    Wenjie Wang\textsuperscript{\rm 1},
    Fuli Feng\textsuperscript{\rm 1}
}
\affiliations {
    \textsuperscript{\rm 1}University of Science and Technology of China\\
    \textsuperscript{\rm 2}National University of Singapore\\
    \{qqqqqzheng, yiyanxu24, chongming.gao, xylin1028, wenjiewang96, fulifeng93\}@gmail.com
}

\begin{document}

\maketitle

\begin{abstract}
With the rapid and continuous increase in academic publications, identifying high-quality research has become an increasingly pressing challenge. While recent methods leveraging Large Language Models (LLMs) for automated paper evaluation have shown great promise, they are often constrained by outdated domain knowledge and limited reasoning capabilities. In this work, we present \textbf{\textit{PaperEval}}, a novel LLM-based framework for automated paper evaluation that addresses these limitations through two key components: 1) a domain-aware paper retrieval module that retrieves relevant concurrent work to support contextualized assessments of novelty and contributions, and 2) a latent reasoning mechanism that enables deep understanding of complex motivations and methodologies, along with comprehensive comparison against concurrently related work, to support more accurate and reliable evaluation. 
To guide the reasoning process, we introduce a progressive ranking optimization strategy that encourages the LLM to iteratively refine its predictions with an emphasis on relative comparison. 
Experiments on two datasets demonstrate that PaperEval consistently outperforms existing methods in both academic impact and paper quality evaluation. In addition, we deploy PaperEval in a real-world paper recommendation system for filtering high-quality papers, which has gained strong engagement on social media---amassing over 8,000 subscribers and attracting over 10,000 views for many filtered high-quality papers---demonstrating the practical effectiveness of PaperEval.
\end{abstract}

\section{Introduction}

In recent years, the explosive growth of academic publications has reflected the vitality of the research community, while simultaneously posing a critical challenge: How can researchers efficiently identify high-quality, impactful work to learn effectively and drive innovation?
In this context, the task of automated paper evaluation is becoming increasingly crucial. 
It aims to evaluate paper quality and predict future impact, thereby facilitating the selection of high-quality work, supporting researchers in navigating the expanding scientific landscape, and ultimately promoting more efficient and impactful research progress.

Technically, the paper evaluation task aims to analyze the paper features to assess research quality from various dimensions, such as academic impact~\cite{10.1007/s11192-022-04547-8,zhao2024literature} and overall quality~\cite{lin2023automated}. 
Existing studies mainly rely on traditional neural models or Large Language Models (LLMs) for this task:
\begin{itemize}[leftmargin=*]
\item \textbf{Traditional methods} utilize neural models, such as Multi-Layer Perceptrons (MLPs), or Long-Short Term Memory networks (LSTM), to evaluate research papers based on predefined features, including structural indicators like paper length and reference count~\cite{vergoulis2020simplifying, Ruan2020PredictingTC}, as well as textual patterns~\cite{ma2021deep, yang2018automatic}. However, these methods often overlook the semantic content of papers (\eg abstract and main text), which is essential for accurate evaluation, ultimately leading to unsatisfactory performance.

\item \textbf{LLM-based methods} leverage rich textual information (\eg title, abstract, and main text) to learn informative paper representations and employ a scoring module to produce evaluation scores. 
Empowered by the advanced semantic understanding capabilities of LLMs, these methods demonstrate strong potential in capturing the technical soundness of research papers, yielding more accurate evaluation results~\cite{lu2024ai, liu2025lmcbert, zhao2025words, de2024can}.
\end{itemize}
Despite promising progress, LLM-based methods still face notable limitations: 1) Due to the time lag in their training data, LLMs often lack awareness of newly published work, making it difficult to compare and assess the novelty and contribution in fast-evolving areas. 
2) Research papers often contain intricate motivations and nuanced methodological designs that require deep reasoning beyond surface-level representation learning. 

To address these limitations, we propose \textbf{\textit{PaperEval}}, 
a framework that retrieves domain-relevant reference papers, jointly encodes them with the target paper into an LLM, and performs latent reasoning to generate accurate evaluation. 
\begin{itemize}[leftmargin=*]
\item \textbf{Domain-aware paper retrieval.} To mitigate the issue of outdated domain knowledge, PaperEval integrates a retrieval module that identifies concurrent and thematically relevant work as reference papers, which are jointly fed into the LLM along with the target paper for evaluation and provide essential context and background for LLMs to better evaluate the novelty and contributions of the target paper within the current research landscape.
\item \textbf{Reasoning-enhanced paper evaluation.}
Evaluating research papers requires a deep understanding of complex motivations and nuanced methodologies. This challenge is further intensified by the need to compare concurrently retrieved work. 
This motivates us to stimulate the reasoning mechanism of LLMs to support deep comprehension, precise comparison, and fair evaluation. 
While chain-of-thought reasoning~\cite{wei2022chain} offers interpretable intermediate steps, it typically requires annotated reasoning paths for supervision~\cite{weng2023large}, which are scarce in paper evaluation scenarios. In contrast, latent reasoning~\cite{hao2024training} enables implicit multi-step reasoning within the hidden representations of LLMs, eliminating the need for explicit annotations. More importantly, LLM-based paper evaluation focuses on learning more informative paper representations to enhance evaluation accuracy, which aligns naturally with the latent reasoning paradigm, seamlessly integrating reasoning directly at the representation level.
As such, we incorporate latent reasoning into PaperEval for comprehensive representation learning.
\end{itemize}

Despite the significant potential of latent reasoning, an effective optimization strategy is essential to guide the reasoning process toward the ultimate ranking goal of paper evaluation. 
Specifically, the paper evaluation task focuses on comparing and identifying valuable work within a large collection, 
with a primary emphasis on relative ranking rather than absolute scoring. 
However, learning accurate rankings is inherently more challenging, as even minor prediction errors may cause substantial shifts in the ranking positions.
To address this, we propose a \textbf{progressive ranking optimization} strategy, which encourages the latent reasoning to progressively improve relative ranking.
In particular, at each reasoning step, we compute the temperature-controlled softmax over the predicted scores of a batch of papers, which is then aligned with the ground-truth order using a listwise ranking loss. 
To gradually refine the LLM's ranking, we progressively decrease the temperature during latent reasoning, making the predicted distributions increasingly sharper and more sensitive to ranking errors. 
This progressive refinement encourages the LLM to iteratively produce more confident and discriminative rankings within each training batch. 
By learning to distinguish fine-grained differences among batch samples, the LLM enhances ranking reasoning capabilities, which can naturally generalize to global ranking across the entire dataset, as theoretically supported by~\cite{10.1145/1553374.1553449}.

We evaluate the effectiveness of PaperEval on two datasets, covering key evaluation dimensions including academic impact and overall quality. 
Extensive experimental results demonstrate its superiority over traditional and LLM-based baselines. 
Furthermore, we deploy PaperEval in a real-world recommendation system to filter high-quality papers from thousands of daily publications. The system powers social media services with over 8,000 subscribers, and several recommended papers have received over 10,000 views on social platforms, demonstrating the practical evaluation effectiveness of PaperEval. 
Our code and data are available in the Supplementary Materials.

In summary, our key contributions are as follows:
\begin{itemize}[leftmargin=*]
\item We propose PaperEval, a novel LLM-based framework for automated paper evaluation that combines a domain-aware paper retrieval module with a latent reasoning mechanism to enable more accurate assessments.
\item We develop a progressive ranking optimization strategy that supervises the LLM reasoning process to iteratively refine its ranking predictions, effectively aligning with the relative ranking objective of paper evaluation.

\item PaperEval achieves state-of-the-art performance on two datasets in both academic impact and paper quality evaluation, demonstrating the superiority of PaperEval with progressive ranking optimization.

\item We deploy PaperEval in a real-world paper recommendation system, which selects the top 10 high-quality papers each day from thousands of new submissions in fast-evolving research areas.

\end{itemize}

\section{Related Work}

\subsection{Paper Evaluation}
Paper evaluation aims to assess a paper’s quality or predict its academic impact. From the quality perspective, a central task is paper rating — predicting whether a paper will be accepted by peer review committees~\cite{lin2023automated}. Existing methods fall into three main categories. The first leverages neural architectures like CNNs and attention-based models to capture local and global textual interactions~\cite{yang2018automatic, deng2020hierarchical}. The second extracts metadata features and employs traditional models such as random forests~\cite{wang2024content}. The third directly encodes the textual content using pretrained models like BERT~\cite{devlin2019bert}, as in recent work~\cite{xue2023re, liu2025lmcbert}.
For evaluating academic impact, prior studies focus on predicting citation counts, citation levels, or other derived impact metrics~\cite{zhao2024literature}, which can similarly be grouped into three strategies. Metadata-based approaches use handcrafted or extracted features with classical models like MLPs or decision trees~\cite{wang2011mining, qiu2024early, Ruan2020PredictingTC, zhang2024predicting}. Graph-based methods model early citation dynamics using citation graphs and apply graph neural networks for future trend prediction~\cite{yan2024modeling, he2023h2cgl, li2023simplifying, jiang2021hints}. LLM-based approaches either prompt models with a paper’s title and abstract to generate citation scores~\cite{de2024can}, or map dense representations to impact-related metrics~\cite{zhao2025words}.
Despite these advances, accurate and fine-grained paper evaluation remains challenging. In this work, we explore how integrating retrieval techniques with the reasoning capabilities of LLMs can address this challenge.

\subsection{Latent Reasoning}
Unlike Chain-of-Thought (CoT) reasoning~\cite{wei2022chain, su2025explicit, carrow2025neural}, which relies on explicitly generated intermediate steps, latent reasoning performs inference directly within a model’s hidden representations~\cite{hao2024training, biran2024hopping}. This implicit approach has gained momentum in LLM-based recommendation~\cite{tang2025think, shen2025efficient, shen2025codi, liu2025lares} and retrieval tasks~\cite{ji2025learning}, as it avoids the need for annotated reasoning traces while enabling richer, more informative representations.
However, a key challenge remains: how to effectively supervise the latent reasoning process. On the one hand, it is essential to ensure that the reasoning process unfolds progressively toward the correct output, allowing the model to refine its prediction step by step~\cite{tang2025think, liu2025lares, ji2025learning}. On the other hand, models must avoid degenerate reasoning, where the hidden states prematurely converge and hinder iterative refinement. For instance, Tang et al.~\cite{tang2025think} introduce a loss that encourages representational diversity across reasoning steps to mitigate this issue.
Despite these efforts, ensuring that the model consistently refines its predictions toward accurate outcomes remains challenging. In this work, we propose a progressive optimization strategy that progressively adjusts the softmax temperature and incorporates a ranking-based loss. This design explicitly guides the latent reasoning process toward better alignment with ground-truth evaluation targets.

\section{PaperEval}
\begin{figure}[t]
\centering
\includegraphics[width=1.0\linewidth]{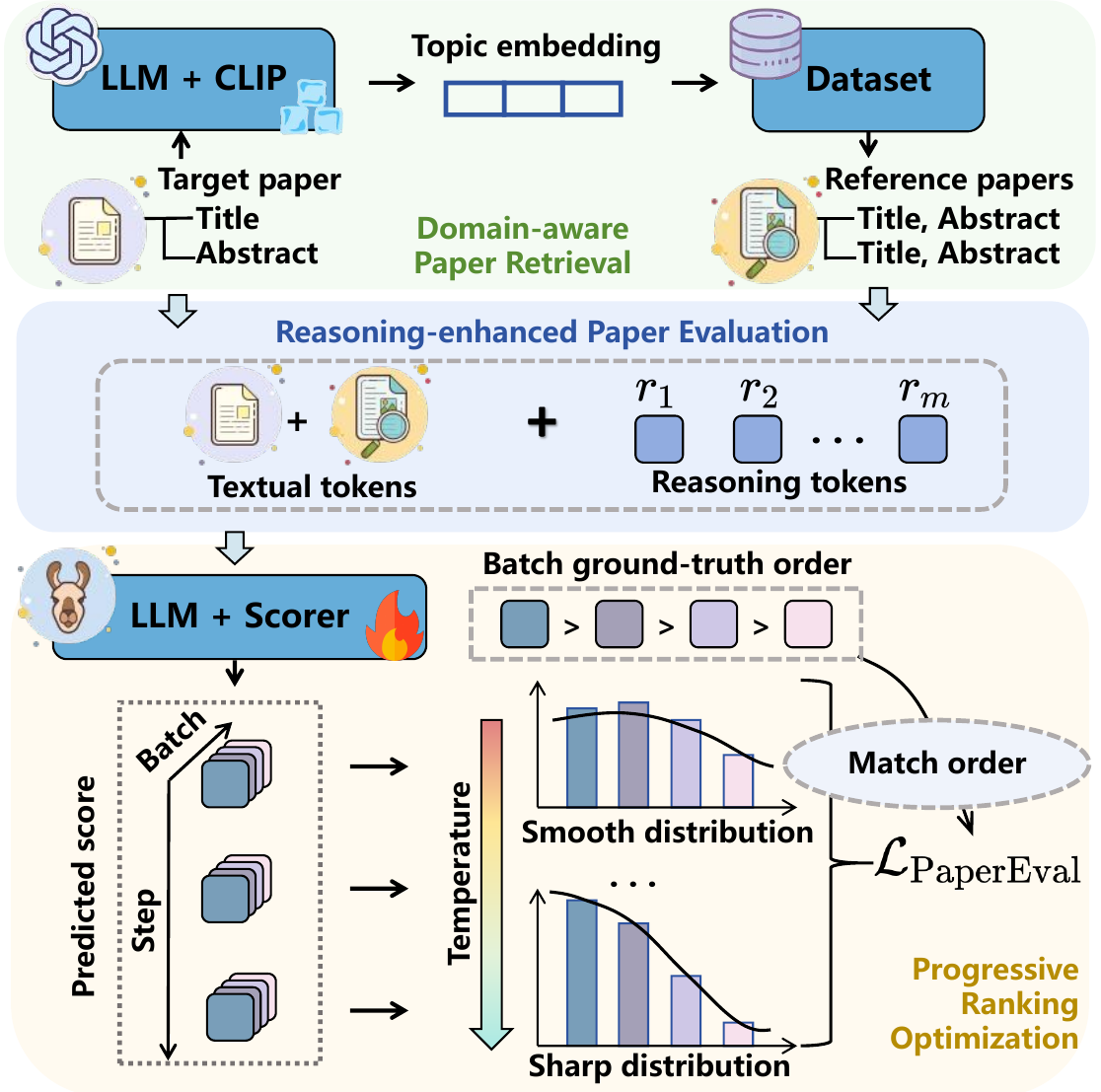}
\caption{Overview of PaperEval. It comprises two key components: domain-aware paper retrieval and reasoning-enhanced paper evaluation, where the model is fine-tuned using the progressive ranking optimization strategy.}
\label{fig:PaperEval}  
\end{figure}


As shown in Figure~\ref{fig:PaperEval}, \textbf{Domain-Aware Paper Retrieval} module selects relevant reference papers, which are jointly encoded with the target paper by \textbf{Reasoning-Enhanced Paper Evaluation} through multi-step latent reasoning to produce a quality prediction. To guide the reasoning toward more accurate ranking, we apply a \textbf{Progressive Ranking Optimization} strategy.

\subsection{LLM-based Paper Evaluation}
Given a set of $N$ research papers $\mathcal{P} = \{p_i\}_{i=1}^N$, where each paper $p_i$ is associated with a ground-truth score $s_i$ reflecting various evaluation aspects (\eg academic impact and overal quality), LLM-based paper evaluation aims to learn informative paper representations that enable accurate prediction of these scores. Due to the high computational cost of processing full paper bodies, recent methods typically utilize the most representative textual elements (e.g., title and abstract) to construct paper representation $w_i$, then integrate a lightweight scorer to produce a predicted score $\hat{s}_i$. Formally,
\begin{equation}
\label{equ:task_pre}
    w_i = \text{LLM}(p_i)[-1],\quad \hat{s}_i = \text{Scorer}(w_i),
\end{equation}
where $\text{LLM}(p_i)[-1]$ denotes the final hidden state output by the LLM for paper $p_i$, and $\text{Scorer}(\cdot)$ is usually implemented as a lightweight MLP.

\subsection{Domain-aware Paper Retrieval}
\label{sec:RAPR}
To equip LLMs with up-to-date domain knowledge for more accurate assessment of the novelty and contributions of each target paper, we introduce a domain-aware retrieval module (depicted at the top of Figure~\ref{fig:PaperEval}). Specifically, given the title and abstract of research papers, we first employ ChatGPT~\cite{achiam2023gpt} to generate representative topic keyphrases, which are then encoded into topic embeddings using the CLIP text encoder~\cite{radford2021learning}. To identify relevant work for each target paper $p_i$, we first compute the cosine similarity between the target topic embedding and those of all other papers in the corpus, where papers with similarity exceeding a predefined threshold $\gamma$ are retained as candidates. Considering the rapidly evolving nature of many research fields, we further filter these candidates by selecting only concurrent relevant papers, whose publication dates are closest to the target paper $p_i$. The resulting set, denoted as $\mathcal{R}_i$, contains at most $k$ papers and serves as the domain-aware reference set, providing essential contextual background to help the LLM more accurately assess the target paper's position within the current research landscape.

\subsection{Reasoning-enhanced Paper Evaluation}
\label{sec:LPR}
To achieve a comprehensive understanding of the motivation and methodological designs of the target paper and effectively incorporate the retrieved domain-aware reference set for contextualized evaluation, PaperEval adopts a latent reasoning mechanism that performs implicit multi-step reasoning for more accurate and reliable evaluation. 

Formally, given the target paper $p_i$ and the corresponding domain-aware reference set $\mathcal{R}_i$, we first construct a textual prompt based on the title and abstract of both target and reference papers, and then tokenize it into a sequence of tokens $T_i$. To stimulate the reasoning capabilities of LLMs (as shown in the middle of Figure~\ref{fig:PaperEval}), PaperEval introduces $m$ reasoning tokens $\{r_1, r_2, \cdots, r_m\}$, which represent intermediate reasoning steps. These tokens guide LLMs to progressively refine the latent states, yielding increasingly informative and discriminative paper representations. The process is formulated as follows:
\begin{equation}
\small
w^{(1)}_i, w_i^{(2)}, \cdots, w_i^{(m)} = \text{LLM}(T_i, r_1, r_2, \cdots, r_m)[-m:],
\end{equation}
where $w_i^{(j)}$ denotes the intermediate paper representation at the $j$-th reasoning step. Each representation is then passed through a scorer to obtain a predicted score: $\hat{s}_i^{(j)} = \text{Scorer}(w_i^{(j)})$, resulting in $m$ predictions, all of which are supervised during training, as detailed in the next section

\subsection{Progressive Ranking Optimization}
Since paper evaluation focuses on identifying the most valuable papers, relative ranking is prioritized over predicting absolute scores. However, learning accurate rankings is challenging, as even small mistakes can lead to substantial shifts in order. To address this, we propose a \textbf{progressive ranking optimization} strategy (illustrated at the bottom of Figure~\ref{fig:PaperEval}) that encourages the model to iteratively refine its predictions during multi-step reasoning, gradually improving its ranking accuracy.

\subsubsection{Training.}

Inspired by ListMLE~\cite{xia2008listwise}, which learns to predict rankings by maximizing the likelihood of the ground-truth order, we adapt it to the paper evaluation scenario, encouraging the predicted scores to yield a ranking consistent with the ground-truth.

Given a training batch of $B$ target papers, we first sort the papers in descending order of their ground-truth scores, which serves as the supervision signal to guide the model to focus on relative rankings. At each reasoning step $j$, the model predicts a batch of scores $\{\hat{s}_i^{(j)}\}_{i=1}^B$, which is converted into a score distribution via a softmax function. As the reasoning process deepens, we hope the model predictions are progressively refined, gradually converging toward the optimal relative rankings with increasing confidence. Motivated by this, we introduce progressive temperature annealing into the softmax function to progressively sharpen the score distribution, which increases confidence in the prediction and amplifies the penalty for incorrect rankings, providing stronger supervision. Specifically, we apply a linearly decreasing temperature schedule:
\begin{equation}
    \tau^{(j)}=\tau_{\max}+\dfrac{j}{m}(\tau_{\min} - \tau_{\max}),
\end{equation}
where $\tau_{\min}<\tau_{\max}\in \mathbb{R}^+$ indicates the upper and lower bounds of temperature, and $\tau^{(j)}$ refers to the annealed temperature for reasoning step $j$. Therefore, the score distribution at step $j$ is defined as:
\begin{equation}
\label{equ:t-distribution}
    \hat{f}^{(j)}_i = \frac{\exp(\hat{s}^{(j)}_i / \tau^{(j)})}{\sum_{t=1}^B \exp(\hat{s}^{(j)}_t / \tau^{(j)})}.
\end{equation}

Based on the score distribution at each reasoning step, we adapt the ListMLE loss into the paper evaluation setting, encouraging the model to generate correct relative rankings:
\begin{equation}
    \mathcal{L} = -\sum_{j=1}^m \log \prod_{i=1}^B \frac{\hat{f}_{r(i)}^{(j)}}{\sum_{k=i}^B \hat{f}_{r(k)}^{(j)}},
\end{equation}
where $r(i)$ denotes the paper ranked at the $i$-th position in the ground-truth permutation. This loss can be interpreted as the log-probability of sequentially sampling papers according to the ground-truth order without replacement, based on the model's predicted scores. It effectively guides the model to progressively refine the score predictions that better reflect the desired relative rankings.


\subsubsection{Inference.} After the multi-step reasoning process, the prediction at the final reasoning step $\hat{s}_i^{(m)}$ represents the most informed and discriminative evaluation of the target paper $p_i$, as it integrates progressively refined judgments across all reasoning steps. Therefore, during inference, we directly adopt $\hat{s}_i^{(m)}$ as the final evaluation score for $p_i$.

\section{Practical Application of PaperEval}

\begin{figure}
    \centering
    \includegraphics[width=1.0\linewidth]{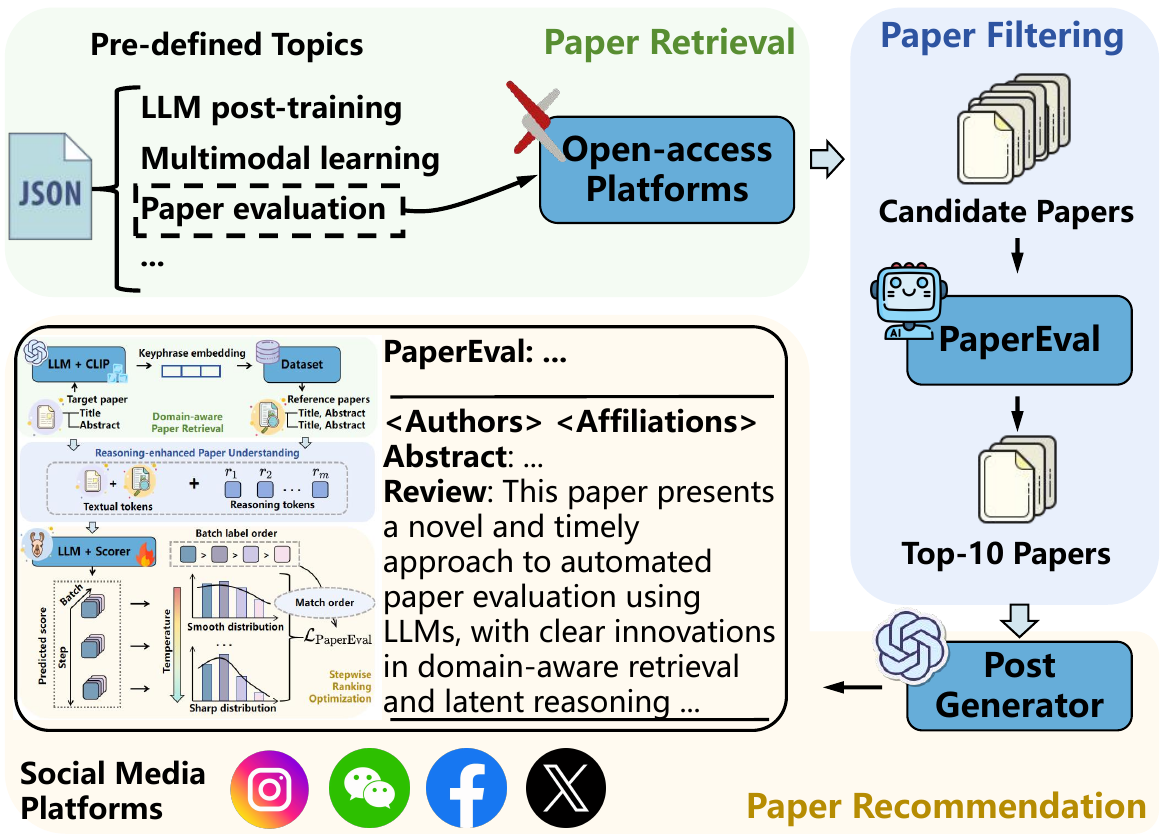}
    \caption{Workflow of the paper recommendation system, which consists of three main phases: paper retrieval, paper filtering, and paper recommendation.}
    \label{fig:PaperRec}
\end{figure}

We deploy \textbf{PaperEval} as the core filtering module of an automated paper recommendation system, which has been successfully launched on social media. As shown in Figure~\ref{fig:PaperRec}, the system operates through three main phases:
\begin{itemize}[leftmargin=*]
\item \textbf{Paper retrieval.} The paper recommendation system first selects key topics from pre-defined topics based on certain rules, and retrieves thousands of candidate papers from open-access paper platforms such as arXiv.
\item \textbf{Paper filtering.} Built upon PaperEval, the system evaluates candidate papers and selects the top 10 most valuable ones based on quality and impact.
\item \textbf{Paper recommendation.} For the selected papers, the system employs a post generator to synthesize concise reviews and organize key information into easily digestible posts, which are then published on social media to provide followers with timely paper recommendations.
\end{itemize}

\section{Experiments}
In this section, we evaluate our proposed PaperEval on two datasets targeting future impact and overall paper quality. We aim to answer the following research questions:

\begin{itemize}[leftmargin=*]
\item \textbf{RQ1:} How does PaperEval compare to both LLM-based models and traditional neural baselines?
\item \textbf{RQ2:} What is the contribution of different components (e.g., paper retrieval, latent reasoning) to the overall performance of PaperEval?
\item \textbf{RQ3:} How does PaperEval improve its ranking predictions through progressive latent reasoning?
\end{itemize}

\subsection{Experimental Settings}
\begin{table*}[t]
\centering
\begin{tabular}{@{}l|cccc|cccc@{}}
\toprule
\multirow{2}{*}{\textbf{Metrics}} & \multicolumn{4}{c|}{\textbf{NAID}}                                                           & \multicolumn{4}{c}{\textbf{ICLR}}                                                            \\
                                  & \textbf{N@10 $\uparrow$} & \textbf{N@20 $\uparrow$} & \textbf{Spearman $\uparrow$} & \textbf{Kendall $\uparrow$} & \textbf{N@10 $\uparrow$} & \textbf{N@20 $\uparrow$} & \textbf{Spearman $\uparrow$} & \textbf{Kendall $\uparrow$} \\ \midrule
\textbf{MLP-based}                      & 0.5109             & 0.5605             & 0.0505                  & 0.2868                   & -                  & -                  & -                       & -                        \\
\textbf{LSTM-based}                     & 0.4506             & 0.4512             & -0.0009                 & 0.0904                   & 0.5119             & 0.5515             & 0.1355                  & 0.0929                   \\
\textbf{GPT-part}                 & 0.5332             & 0.5258             & 0.0748                  & 0.0527                   & 0.6600             & 0.6428             & 0.0572                  & 0.0417                   \\
\textbf{GPT-all}                  & -                  & -                  & -                       & -                        & 0.6268             & 0.6365             & 0.0579                  & 0.0481                   \\
\textbf{SciBERT}                  & 0.5784             & 0.5615             & 0.0365                  & 0.2709                   & 0.6491             & 0.6877             & 0.2114                  & 0.1457                   \\
\textbf{NAIP}                     & 0.9274             & 0.9079             & 0.4514                  & 0.3163                   & 0.7510             & 0.7306             & 0.3188                  & 0.2236                   \\ \midrule
\textbf{PaperEval}         & \textbf{0.9589}    & \textbf{0.9521}    & \textbf{0.4953}         & \textbf{0.3438}          & \textbf{0.7784}    & \textbf{0.7386}    & \textbf{0.3276}         & \textbf{0.2285}          \\
\bottomrule
\end{tabular}
\caption{Performance comparison of PaperEval and baseline models on the NAID and ICLR datasets.
N@\{10,20\} denotes NDCG@\{10,20\}, while Spearman and Kendall represent Spearman's rho and Kendall's tau, respectively. The best performance for each metric is shown in \textbf{bold}.}
\label{table:performance}
\end{table*}

\subsubsection{Datasets.}

To evaluate the performance of PaperEval, we conduct experiments on two datasets. The NAID dataset, which is publicly available, provides scores reflecting scientific impact. Furthermore, we construct a new dataset, the ICLR-based dataset, to assess research quality through peer review scores. This dual perspective allows for a comprehensive evaluation of our model's ability to predict both long-term scholarly influence and immediate research quality. 

(1) \textbf{NAID}~\cite{zhao2025words}: This dataset is derived from arXiv. Each paper includes its title, abstract, some metadata (e.g., paper length, number of references), and an impact score that quantifies its relative citation rank within the same domain and publication period, serving as an indicator of the paper’s future impact in its field. The dataset contains 11,118 papers in the training set and 1,237 papers in the test set. 

(2) \textbf{ICLR}: This dataset contains peer review data from ICLR 2021 to 2024 via the OpenReview platform. Each paper includes the title and abstract. To compute paper ratings, we first calculate the average of all review scores, remove scores that deviate by more than 3 points from this average, and then compute the average of the remaining scores as the final scores. This score is normalized to the range $[0, 1]$ and used as the overall quality score. The training set consists of 14,914 papers, and the test set consists of 1657 papers.

For both datasets, we randomly split 10\% training data as the validation dataset during training.

\subsubsection{Baselines.}
We compare PaperEval against various baselines, including both traditional and LLM-based methods:
\begin{itemize}[leftmargin=*]
\item \textbf{Traditional methods:} 1) \textbf{MLP-based}~\cite{Ruan2020PredictingTC} method uses metadata of the target paper as input to an MLP to predict the evaluation score. We do not evaluate this method on the ICLR dataset due to the lack of metadata. 2) \textbf{LSTM-based}~\cite{ma2021deep} method encodes the abstract of each paper into a sequence representation and applies an LSTM network to predict the target score.
\item \textbf{LLM-based methods:}  3) \textbf{GPT-part}~\cite{de2024can} method prompts ChatGPT to predict a paper’s score based solely on its title and abstract. We use GPT-4o~\cite{hurst2024gpt} as the underlying model. 4) \textbf{GPT-all}~\cite{lu2024ai} approach treats the LLM as a reviewer, prompting it to read the entire paper and generate a full review, including an overall quality score. For cost considerations, we choose GPT-4o-mini as our base model. 5) \textbf{SciBERT}~\cite{beltagy2019scibert} method is a BERT-based model pretrained on scientific text. We fine-tune it with a simple regression module to perform paper evaluation. 6) \textbf{NAIP}~\cite{zhao2025words} method uses LLaMA3-Smaug~\cite{pal2024smaug} as the backbone LLM. An additional regression module is applied to the output embedding to generate the final score.
\end{itemize}

\subsubsection{Evaluation Metrics.}
We adopt a variety of evaluation metrics targeting two key aspects: top-K ranking quality and overall ranking consistency. Rankings are computed over all test samples based on their predicted scores. For detailed definitions, please refer to Supplementary Materials~\ref{appendix:evaluation_metrics}.

\begin{itemize}[leftmargin=*]
\item \textbf{Top-K ranking quality}: Following~\cite{zhao2025words}, we use NDCG@\{10,20\} to measure the ranking quality of the top-K recommended papers.
\item \textbf{Overall ranking consistency}: Following~\cite{Ng2015BetterSE}, we employ Spearman’s rho and Kendall’s tau to assess how well the predicted rankings align with the ground-truth rankings.
\end{itemize}


\subsection{Overall Performance (RQ1)}
Table~\ref{table:performance} presents a comprehensive comparison between PaperEval and all baseline methods. We summarize our key observations as follows:

\begin{itemize}[leftmargin=*]
\item 
MLP-based models that leverage metadata outperform LSTM baselines on the NAID dataset, highlighting the value of metadata features. However, their lack of semantic understanding limits their overall evaluation capability.

\item Pretrained LLMs significantly outperform MLP-based, LSTM-based baselines. This highlights the effectiveness of language model–based semantic understanding in evaluating research papers. Moreover, fine-tuned models (SciBERT, NAIP) achieve better performance than prompting approaches (GPT-part, GPT-all), indicating that task-specific fine-tuning can more effectively enhance a model’s capability in evaluating scientific papers. Furthermore, NAIP outperforms the SciBERT-based method, suggesting that large-scale language models possess stronger semantic capabilities for understanding scientific papers, thereby achieving better performance.

\item PaperEval consistently achieves \textbf{state-of-the-art} performance across all evaluation metrics and datasets. By retrieving domain-relevant references and employing latent reasoning to model complex academic semantics, PaperEval more accurately captures the novelty and quality of target papers, resulting in superior performance in both top-K ranking quality and overall ranking consistency.

\end{itemize}

\subsection{In-depth Analysis (RQ2)}
\setlength{\tabcolsep}{4pt}
\begin{table}[t]
\centering
\begin{tabular}{l|cc|cc}
\toprule
\multirow{2}{*}{\textbf{Method}} & \multicolumn{2}{c|}{\textbf{NAID}}  & \multicolumn{2}{c}{\textbf{ICLR}}   \\
                                 & \textbf{N@10}   & \textbf{Spearman} & \textbf{N@10}   & \textbf{Spearman} \\ \midrule
\textbf{PaperEval}               & 0.9589 & 0.4953   & 0.7784 & 0.3276   \\
\textbf{- w/o Ret.}               & 0.9432          & 0.4702            & 0.7235          & 0.3545            \\
\textbf{- w/o Rea.}         & 0.9449          & 0.4968            & 0.7559          & 0.3479            \\
\textbf{- w/o Opt.}              & 0.9585          & 0.4808            & 0.7166          & 0.3198            \\ \bottomrule
\end{tabular}
\caption{Effect of designs in PaperEval. ``Ret.'' denotes domain-aware paper retrieval method, ``Rea.'' denotes the latent reasoning progress, ``Opt.'' denotes our proposed progressive ranking optimization.}
\label{table:ablation}
\end{table}

In this section, we conduct experiments to further investigate how the designs in PaperEval affect the performance.

\subsubsection{Ablation Study.}
To assess the contribution of each design in PaperEval, we conduct ablations on the NAID dataset: 1) ``w/o Ret.'' removes the paper retrieval module. 2) ``w/o Rea'' skips reasoning and directly outputs predictions. 3) ``w/o Opt.'' replaces progressive ranking optimization with Mean Squared Error (MSE) on final scores.

From the experimental results shown in Table~\ref{table:ablation}, we observe: 1) The performance decline without domain-aware retrieval underscores the importance of contextual references in enhancing evaluation quality. 2) Latent reasoning brings clear gains in top-k performance by better aligning papers with retrieved papers. However, its tendency to converge quickly, combined with the difficulty of supervising the reasoning process, may slightly hurt overall ranking consistency. 3) Removing the optimization strategy significantly reduces performance, since directly regressing dense relevance scores makes it difficult for the model to distinguish subtle differences in paper quality and relative order.

\subsubsection{Loss Variants Comparison.}
To assess the effectiveness of our list-wise ranking loss design, we conduct a series of experiments comparing it with alternative loss designs, all following the same progressive temperature-controlled setting. Specifically, we evaluate the following variants:
\begin{itemize}[leftmargin=*]
\item \textbf{Pair-wise ranking}: Inspired by RankNet~\cite{burges2005learning}, we design a temperature-controlled pair-wise ranking loss to examine whether pair-wise supervision is more suitable for PaperEval than list-wise ranking.
\item \textbf{Distribution similarity}: To investigate whether aligning the predicted and target distributions is the key factor, we generate the ground-truth distribution via a temperature-controlled softmax and compute the KL-divergence between it and the predicted distribution as the loss.
\item \textbf{Score regression}: As a control setting, similar to~\cite{zhao2025words}, we remove the ranking-based objective and directly apply an MSE loss between the final predicted score and the ground-truth label. This setup allows us to examine whether ranking is a more effective learning signal than direct score supervision.
\end{itemize}

\begin{table}[t]
\centering
\begin{tabular}{l|cc|cc}
\toprule
\multirow{2}{*}{\textbf{Method}} & \multicolumn{2}{c|}{\textbf{NAID}}  & \multicolumn{2}{c}{\textbf{ICLR}}   \\
                                 & \textbf{N@10}   & \textbf{Spearman} & \textbf{N@10}   & \textbf{Spearman} \\ \midrule
\textbf{List-wise}       & 0.9589 & 0.4953   & 0.7784 & 0.3276   \\
\textbf{Pair-wise}       & 0.9296          & 0.4738            & 0.7350          & 0.3139            \\
\textbf{Distribution} & 0.9265          & 0.4772            & 0.7559          & 0.3071            \\
\textbf{Regression}              & 0.9585          & 0.4808            & 0.7166          & 0.3198            \\ \bottomrule
\end{tabular}
\caption{Performance comparison of various loss designs.}
\label{table:loss_design}
\end{table}
\setlength{\tabcolsep}{6pt}
For implementation details of each loss variant, please refer to the Supplementary materials~\ref{appendix:loss_design}. To ensure a fair comparison, we perform hyperparameter tuning (learning rate, number of epochs) for each variant. 

From the results summarized in Table~\ref{table:loss_design}, we observe the following:
1) List-wise ranking loss outperforms pair-wise ranking loss.
This suggests that optimizing over the full ranking list provides a more informative and fine-grained training signal than relying solely on pairwise comparisons. The list-wise objective encourages the model to consider global ranking consistency, which is particularly beneficial in complex evaluation tasks like ours.
2) The distribution similarity loss fails to capture fine-grained relative order. While it encourages the overall score distribution to match the target distribution, it lacks explicit supervision over the relative ranking between individual papers. As a result, its performance falls short, indicating that alignment at the distribution level is insufficient for our goal of precise paper ranking.
3) Unlike ranking-based objectives, score regression focuses on predicting exact values, which often misaligns with identifying the relative ranking of the papers. As our results show, regression consistently underperforms on all metrics, confirming that optimizing for relative order is more suitable and effective.

\subsubsection{Hyperparameter Analysis.}
\begin{figure}[t]
    \centering
    \includegraphics[width=1.0\linewidth]{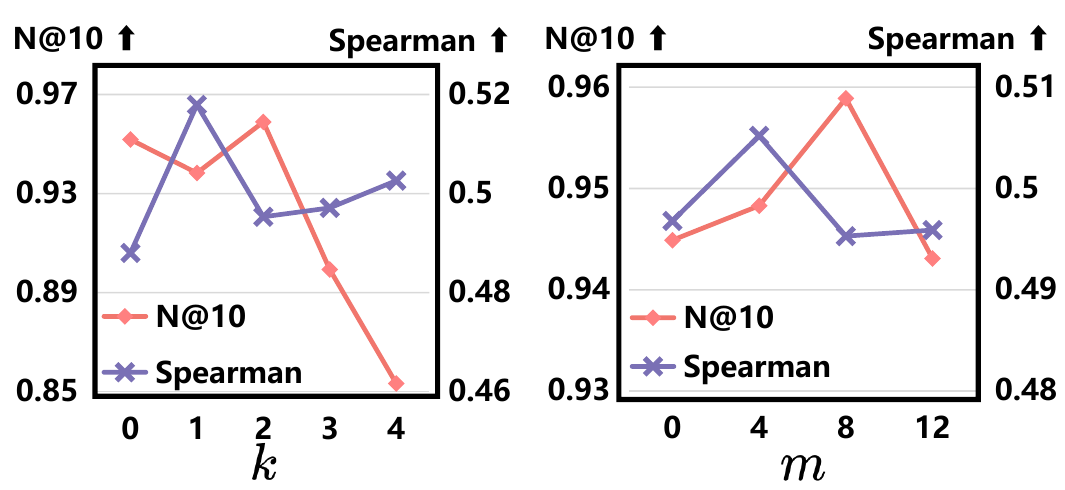}
    \caption{Performance comparison with different hyperparameter settings. We vary the number of reference papers $k$ and the number of reasoning steps $m$.}
    \label{fig:hyperparams}
\end{figure}
In this section, we examine the effectiveness of two key hyperparameters on the NAID dataset: the number of retrieved reference papers $k$ and reasoning steps $m$. The results are shown in Figure~\ref{fig:hyperparams}.

\begin{itemize}[leftmargin=*]
\item \textbf{The number of retrieved reference papers $k$.}
When the number of retrieved reference papers is too small, the LLM lacks sufficient reference context to support accurate evaluation, leading to limited knowledge grounding and poorer ranking performance. Conversely, when too many references are included, the model struggles to effectively capture their relevance to the target paper. This often causes it to lose focus on the target paper itself, impairing top-k identification accuracy. Notably, we observe that NDCG drops more significantly than Spearman in this case, indicating that excessive references particularly affect top-ranked results. Furthermore, a larger $k$ increases input length and computational cost. These findings suggest that selecting a moderate $k$ is essential to balance contextual richness, ranking focus, and efficiency.

\item \textbf{The number of reasoning steps $m$.}
From the right part of Figure~\ref{fig:hyperparams}, we observe a clear trend where performance first increases and then decreases as the number of reasoning steps $m$ grows. On the one hand, using too few reasoning steps fails to fully leverage the reasoning process: the LLM produces an output without sufficiently analyzing or inferring from the input, leading to suboptimal results. On the other hand, using too many reasoning steps also degrades performance. This is primarily due to the increased risk of overfitting. We observe during training that models with large $m$ tend to converge quickly but subsequently overfit severely.
This overfitting can be attributed to the repeated refinement loop in the reasoning process: when the number of steps is excessive, the model repeatedly reprocesses the same input, leading to a kind of memorization or confirmation bias instead of genuine reasoning. Consequently, the model may lose generalizability and begin to reinforce incorrect intermediate conclusions.
Similar to the choice of $k$, it is crucial to select a moderate number of reasoning steps. A well-balanced $m$ encourages the model to reason carefully and refine its predictions while avoiding the pitfalls of excessive iteration.
\end{itemize}

\subsubsection{Analysis of backbone LLMs.} Due to space limits, detailed analysis is provided in the Supplementary Materials.

\subsection{Effect of Latent Reasoning (RQ3)}

\begin{figure}[t]
\centering
\includegraphics[width=1.0\linewidth]{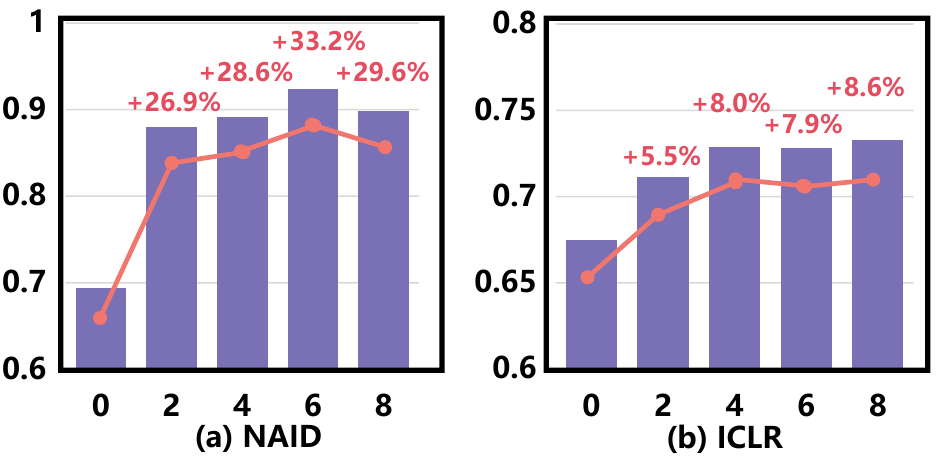}
\caption{N@10 performance across reasoning steps to evaluate progressive refinement.}
\label{fig:steps}
\end{figure}
To assess whether our progressive ranking optimization effectively encourages step-by-step refinement, we analyze model outputs at different reasoning steps. 

We analyze a partially trained model (before full convergence), where step-wise refinement is more visible. After convergence, due to supervision at each step, predictions tend to stabilize, making refinement less apparent. We report NDCG@10 for its sensitivity to ranking quality. “Step-0” denotes the baseline prediction without reasoning, using the final token output directly.

As shown in Figure~\ref{fig:steps}, the model progressively improves its predictions across reasoning steps, validating the effectiveness of our proposed strategy. The NDCG score consistently rises, reflecting improved alignment with the ground-truth ranking. However, in Figure~\ref{fig:steps}(a), a slight drop at the final step suggests that prolonged latent reasoning may lead to diminishing returns or repetitive thinking, pointing to a potential limitation and direction for future work.

\section{Conclusion}


In this work, we focus on automatic paper evaluation, which involves assessing specific aspects of research papers to help researchers navigate the growing volume of academic publications. We propose PaperEval, a novel LLM-based framework that combines domain-aware retrieval with latent reasoning to enable more accurate and reliable evaluations. Furthermore, we design a progressive ranking optimization strategy that guides the reasoning process by progressively refining predictions toward more accurate relative rankings. Experimental results demonstrate that our framework achieves state-of-the-art performance in both academic impact and overall quality assessment. Besides, we deploy PaperEval in a real-world paper recommendation system, which has gained notable traction on social media, attracting over 8,000 subscribers and generating more than 10,000 views for several recommended papers.

Despite strong performance, our framework still has limitations, particularly for the latent reasoning module, which opens promising directions for future research. Enhancing latent reasoning for more accurate and insightful paper evaluation requires more effective supervision strategies that are robust to hyperparameter choices, computationally efficient, and resilient to overfitting. In addition, incorporating multimodal information (\eg figures and tables) from research papers presents another valuable avenue for enhancing the accuracy and depth of paper evaluation.

\bibliography{refer.bib}

\section{Reproducibility Checklist}
This paper:
\begin{itemize}
\item Includes a conceptual outline and/or pseudocode description of AI methods introduced. \textbf{Yes.}
\item Clearly delineates statements that are opinions, hypothesis, and speculation from objective facts and results. \textbf{Yes.}
\item Provides well marked pedagogical references for less-familiare readers to gain background necessary to replicate the paper. \textbf{Yes.}
\end{itemize}
Does this paper make theoretical contributions? \textbf{No.} \\
Does this paper rely on one or more datasets? \textbf{Yes.} 
\begin{itemize}
\item A motivation is given for why the experiments are conducted on the selected datasets. \textbf{Yes.}
\item All novel datasets introduced in this paper are included in a data appendix. \textbf{Yes.}
\item All novel datasets introduced in this paper will be made publicly available upon publication of the paper with a license that allows free usage for research purposes. \textbf{Yes.}
\item All datasets drawn from the existing literature (potentially including authors’ own previously published work) are accompanied by appropriate citations. \textbf{Yes.}
\item All datasets drawn from the existing literature (potentially including authors’ own previously published work) are publicly available. \textbf{Yes.}
\item All datasets that are not publicly available are described in detail, with explanation why publicly available alternatives are not scientifically satisficing. \textbf{NA.}
\end{itemize}
Does this paper include computational experiments? \textbf{Yes.}
\begin{itemize}
\item This paper states the number and range of values tried per (hyper-) parameter during development of the paper, along with the criterion used for selecting the final parameter setting. \textbf{Yes.}
\item Any code required for pre-processing data is included in the appendix. \textbf{Yes.}
\item All source code required for conducting and analyzing the experiments is included in a code appendix. \textbf{Yes.}
\item All source code required for conducting and analyzing the experiments will be made publicly available upon publication of the paper with a license that allows free usage for research purposes. \textbf{Yes.}
\item All source code implementing new methods have comments detailing the implementation, with references to the paper where each step comes from. \textbf{Yes.}
\item If an algorithm depends on randomness, then the method used for setting seeds is described in a way sufficient to allow replication of results. \textbf{Yes.}
\item This paper specifies the computing infrastructure used for running experiments (hardware and software), including GPU/CPU models; amount of memory; operating system; names and versions of relevant software libraries and frameworks. \textbf{Partial.}
\item This paper formally describes evaluation metrics used and explains the motivation for choosing these metrics. \textbf{Yes.}
\item This paper states the number of algorithm runs used to compute each reported result. \textbf{No.}
\item Analysis of experiments goes beyond single-dimensional summaries of performance (e.g., average; median) to include measures of variation, confidence, or other distributional information. \textbf{No.}
\item The significance of any improvement or decrease in performance is judged using appropriate statistical tests (e.g., Wilcoxon signed-rank). \textbf{No.}
\item This paper lists all final (hyper-)parameters used for each model/algorithm in the paper’s experiments. \textbf{Partial.}
\end{itemize}

\clearpage
\newpage
\appendix
\section{Supplementary Material}

This supplementary material provides additional details on the implementation, evaluation, and experimental designs of PaperEval. We include extended explanations of the evaluation metrics, loss function comparisons, and in-depth effect analysis of batch size and backbone LLM. Furthermore, we present all the prompts used in the experiments and release all the code and datasets to facilitate reproducibility:

\begin{links}
\link{Code and Datasets}{https://github.com/ZhengWwwq/PaperEval}
\end{links}


\subsection{Implementation Details.}
Following NAIP~\cite{zhao2025words}, we adopt LLaMA3-Smaug~\cite{pal2024smaug} as the backbone of PaperEval for fair comparison. We set the number of reasoning steps $m$ to 8 and the number of retrieved reference papers $k$ to 2. We set the temperature $\tau_{\text{max}}$ to $1.0$ and $\tau_{\text{min}}$ to $0.1$. We train the model for 5 epochs on both the NAID and ICLR datasets with a shared learning rate of $5.0 \times 10^{-5}$, and select the checkpoint with the best validation performance for evaluation. All experiments are conducted on a single NVIDIA A40 GPU.

\subsection{Evaluation Metrics}
\label{appendix:evaluation_metrics}
To quantitatively assess the quality of ranked outputs in our experiments, we adopt three widely used evaluation metrics: Normalized Discounted Cumulative Gain (NDCG), Spearman’s rank correlation coefficient, and Kendall’s rank correlation coefficient. These metrics evaluate how well the predicted rankings align with ground-truth rankings, from both top-weighted and pairwise perspectives.

\subsubsection{Normalized Discounted Cumulative Gain (NDCG).}
NDCG is a position-sensitive metric commonly used in information retrieval to evaluate the relevance of ranked items. Given a list of items ranked by a model, it compares the predicted ranking with the ground-truth ranking, placing more emphasis on correctly ordering higher-ranked items. The DCG for a list of length $K$ is defined as:
\begin{equation}
\mathrm{DCG@K} =\sum_{i=1}^K \frac{2^{\mathrm{rel}_i}-1}{\log _2(i+1)},
\end{equation}
where $rel_i$ is the ground-truth relevance score of the item at position $i$. The NDCG is obtained by normalizing DCG by the ideal DCG (IDCG), which is the DCG of the ground-truth ranking:
\begin{equation}
\text { NDCG@K }=\frac{\text { DCG@K }}{\text { IDCG@K }}.
\end{equation}
NDCG ranges from 0 to 1, with higher values indicating better ranking quality.

\subsubsection{Spearman’s Rank Correlation Coefficient.}
Spearman’s $\rho$ measures the rank correlation between two variables by assessing how well the relationship between the ground-truth and predicted rankings can be described by a monotonic function. It is defined as the Pearson correlation between the ranks of the data:
\begin{equation}
\rho=1-\frac{6 \sum d_i^2}{n\left(n^2-1\right)},
\end{equation}
where $d_i$ is the difference between the predicted and ground-truth ranks of the $i$-th item, and $n$ is the number of items. A $\rho$ value of 1 implies perfect agreement, 0 implies no correlation, and -1 indicates perfect inverse correlation.

\subsubsection{Kendall’s Tau.}
Kendall’s $\tau$ is another rank correlation metric that focuses on the number of concordant and discordant pairs between two rankings. For a set of $n$ items, Kendall’s $\tau$ is computed as:
\begin{equation}
\tau=\frac{C-D}{\frac{1}{2} n(n-1)}, 
\end{equation}
where $C$ is the number of concordant pairs (i.e., item pairs ranked in the same order in both lists) and $D$ is the number of discordant pairs. Like Spearman’s $\rho$, $\tau$ ranges from -1 (complete disagreement) to 1 (complete agreement), with higher values indicating stronger alignment between predicted and ground-truth rankings.

Together, these metrics offer complementary perspectives on ranking performance: NDCG captures the utility of top-ranked items, while Spearman’s $\rho$ and Kendall’s $\tau$ assess the overall consistency of the rank ordering.

\subsection{Loss Function Comparison}
\label{appendix:loss_design}
To investigate which type of loss function is most suitable for training LLMs for paper evaluation, we explore alternative designs: different list-wise ranking losses, a pair-wise ranking loss, and a simple score regression loss. All follow our proposed progressive temperature-controlled training framework. In addition to the comparisons presented in the main text (Table 3), we conduct further experiments on the NAID dataset to explore more variants. These results serve as a supplement to the main analysis. We detail the formulation of each loss function below. Notation is consistent with the main text.

\subsubsection{ListNet Ranking Loss.}
Inspired by~\cite{yang2020enhancing}, we also design a distribution-based similarity/ListNet~\cite{cao2007learning} loss to encourage the predicted score distribution to match the ground-truth score distribution. Specifically, we convert both the predicted scores and the ground-truth scores into soft probability distributions using a temperature-controlled softmax. The temperature varies with the reasoning step to enable coarse-to-fine comparison. The ground-truth distribution at step \( j \) is computed as:

\begin{equation}
    f^{(j)}_i = \frac{\exp(s_i / \tau^{(j)})}{\sum_{t=1}^B \exp(s_t / \tau^{(j)})},
\end{equation}
where \( s_i \) denotes the ground-truth score of paper \( i \), \( B \) is the batch size, and \( \tau^{(j)} \) is the temperature at step \( j \). The predicted distribution \( \hat{f}_i^{(j)} \) is computed in the same way using the predicted scores.

We then compute the Kullback–Leibler (KL) divergence between the ground-truth and predicted distributions at each reasoning step, and sum across all steps to get the final loss:

\begin{equation}
    \mathcal{L}_{\text{ListNet}} = \sum_{j=1}^m \mathbb{D}_{\mathrm{KL}}(f^{(j)} \| \hat{f}^{(j)}).
\end{equation}

\subsubsection{RankCosine Ranking Loss.}
We also try the RankCosine ranking loss~\cite{qin2008query} in PaperEval. RankCosine treats the predicted and ground-truth score lists as vectors and maximizes their cosine similarity, encouraging the predicted ranking to align closely with the target. In addition, we introduce temperature control to progressively sharpen the predicted vector during reasoning steps as follows:

\begin{equation}
    \mathcal{L}_{\text{RankCosine}} = \sum_{j=1}^m \frac{1}{2} (1 - f^{(j)} \odot \hat{f}^{(j)}),
\end{equation}
where the $\odot$ represents the dot between two vectors.

\subsubsection{ApproxNDCG Ranking Loss.}
The Approximate NDCG~\cite{qin2010general} loss is a differentiable surrogate for the standard NDCG metric, enabling its direct optimization in learning-to-rank models. It overcomes the non-differentiability of the ranking operation by replacing the discrete integer rank of an item with a continuous ``soft rank'' $\hat{\pi}_i$. This soft rank is calculated from pairwise comparisons of item scores using a sigmoid function $\sigma$ scaled by a temperature parameter $\tau$.

The final loss function is defined as:
\begin{equation}
  \mathcal{L}_{\text{ApproxNDCG}} = \sum_{j=1}^m(1 - \frac{1}{\text{IDCG}} \sum_{i=1}^{B} \frac{2^{\text{rel}_i} - 1}{\log_2(1 + \hat{\pi}^{(j)}_i)}),
\end{equation}
where the soft rank $\hat{\pi}^{(j)}_i$ for item $i$ is given by:
\begin{equation}
  \hat{\pi}^{(t)}_i = 1 + \sum_{j \neq i} \sigma\left(\frac{s^{(t)}_j - s^{(t)}_i}{\tau^{(t)}}\right).
\end{equation}

\subsubsection{RankNet Ranking Loss.}
We adopt a pair-wise ranking loss inspired by RankNet~\citep{burges2005learning} to guide the model in learning relative paper quality comparisons. Specifically, at each evaluation step $t \in \{1, 2, \cdots, m\}$ we focus on all pairs of papers $(i, j)$ where paper $i$ is preferred over paper $j$, means $s_i > s_j$. The model predicts scalar scores $s_i^{(t)}$ and $s_j^{(t)}$ for the two papers at step $t$, and the loss encourages $s_i^{(t)}$ to be higher than $s_j^{(t)}$ . The pair-wise loss at step $t$ is defined as:
\begin{equation}
    \mathcal{L}_{\text{RankNet}} = - \sum_{t=1}^m\sum_{s_i > s_j} \log \sigma(\frac{\hat{s}_i^{(t)} - \hat{s}_j^{(t)}}{\tau^{(t)}}).
\end{equation}

\subsubsection{Score Regression Loss.}
As a straightforward baseline, we follow previous work and adopt a simple regression objective. This loss aims to directly regress the predicted final score toward the ground-truth label. Specifically, we apply the mean squared error (MSE) between the final predicted score at the last refinement step and the ground-truth score. Formally, the loss is defined as:
\begin{equation}
\mathcal{L}_{\text{MSE}} = \frac{1}{B}\sum_{i=1}^B(\hat{s}_i^{(m)} - s_i)^2,
\end{equation}
where $\hat{s}_i^{(m)}$ denotes the predicted score after the final refinement step $m$, and $s_i$ is the corresponding ground-truth score.

\begin{table}[t]
\centering
\begin{tabular}{lccc}
\toprule
\textbf{Ranking loss} & \textbf{Type} & \textbf{N@10} & \textbf{Spearman} \\ \midrule
\textbf{ListMLE}      & Listwise      & 0.9589             & 0.4953                  \\
\textbf{ListNet}      & Listwise      & 0.9265             & 0.4772                  \\
\textbf{RankCosine}   & Listwise      & 0.9283             & 0.4933                  \\
\textbf{ApproxNDCG}   & Listwise      & 0.8994             & 0.4779                  \\
\textbf{RankNet}      & Pairwise      & 0.9296             & 0.4738                  \\
\textbf{MSE}          & Regression    & 0.9585             & 0.4808                  \\ \bottomrule
\end{tabular}
\caption{Performance comparison of various loss designs on the NAID dataset. In Table 3 in the main text, List-wise refers to ListMLE, and Distribution corresponds to ListNet.}
\label{tab:loss_design_more}
\end{table}

We conduct all loss design experiments on the NAID dataset. From the results in Table~\ref{tab:loss_design_more}, we make two key observations: 1) ListMLE achieves the best performance among all loss functions, as its permutation probability modeling directly reflects the ground-truth ranking, making it particularly well-suited to our training objective. This also supports the theoretical analysis in~\cite{10.1145/1553374.1553449}. 2) Most other listwise and pairwise losses fail to outperform simple MSE, indicating that effective ranking supervision remains challenging compared to direct score prediction.
\setlength{\tabcolsep}{4pt}
\begin{table*}[!h]
\centering
\begin{tabular}{l|c|cccc|cccc}
\toprule
\multirow{2}{*}{\textbf{BaseLLM}} & \multirow{2}{*}{\textbf{Size}} & \multicolumn{4}{c|}{\textbf{NAID}}                    & \multicolumn{4}{c}{\textbf{ICLR}}                     \\
                                  &                                & \textbf{N@10}   & \textbf{N@20}   & \textbf{Spearman} &  \textbf{Kendall} & \textbf{N@10}   & \textbf{N@20}   &   \textbf{Spearman} & \textbf{Kendall} \\ \midrule
\textbf{NAID}                     & 8B                             & 0.9274          & 0.9079          & 0.4514   &  0.3163         & 0.7510          & 0.7306          & 0.3188    & 0.2236        \\ \midrule
\textbf{Llama*}~\cite{pal2024smaug}                   & 8B                             & 0.9589          & \textbf{0.9521} & \textbf{0.4953} & \underline{0.3438}   & \textbf{0.7784} & \underline{0.7386}    & \underline{0.3276} & \underline{0.2285}     \\
\textbf{Llama}~\cite{grattafiori2024llama3herdmodels}                    & 8B                             & \textbf{0.9724} & 0.9038          & 0.4895  & 0.3405          & 0.7276          & 0.7306          & 0.3004  & 0.2091          \\
\textbf{Qwen}~\cite{qwen2025qwen25technicalreport}                     & 7B                             & \underline{0.9691}    & \underline{0.9314}    & \underline{0.4950} & \textbf{0.3446}     & \underline{0.7698}    & \textbf{0.7554} & \textbf{0.3332} & \textbf{0.2320}  \\
\textbf{Mistral}~\cite{jiang2023mistral7b}                  & 7B                             & 0.8260          & 0.8355          & 0.4267   & 0.2944         & 0.7055          & 0.7041          & 0.3055   & 0.2141         \\
\textbf{Phi}~\cite{abdin2024phi3technicalreporthighly}                      & \textbf{4B}                    & 0.7617          & 0.7832          & 0.4864  & 0.3370           & 0.7225          & 0.7180          & 0.2791     & 0.1943       \\ \bottomrule
\end{tabular}
\caption{Performance comparison of various LLMs. ``LLaMA*'' denotes Llama3-Smaug~\citep{pal2024smaug}.}
\label{table:LLM}
\end{table*}
\setlength{\tabcolsep}{6pt}
\subsection{Effect of Batch Size}
\begin{figure}[t]
\centering
\includegraphics[width=1.0\linewidth]{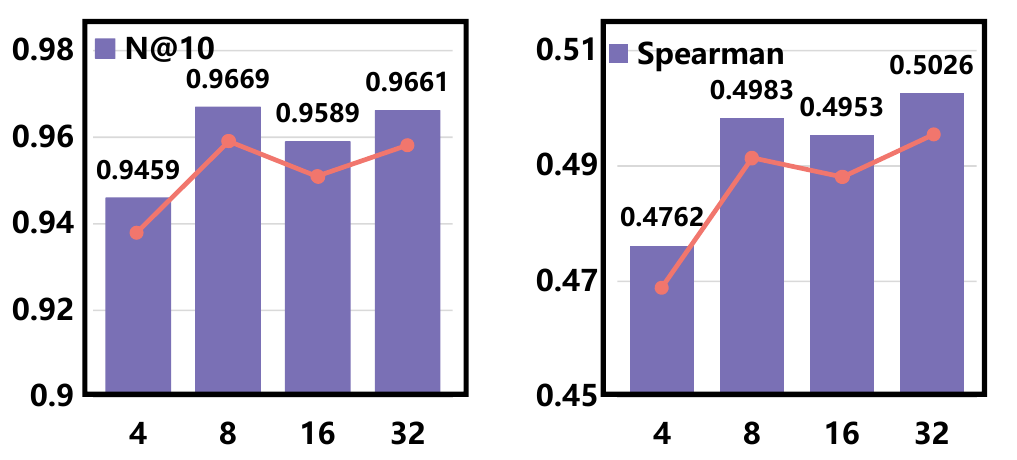}
\caption{Performance comparison with different batch sizes on the NAID dataset.}
\label{fig:bsz}
\end{figure}
As the batch size determines the list length in listwise ranking, we vary it to examine its impact on the performance of our progressive ranking optimization.

The results in Figure~\ref{fig:bsz} show that larger batch sizes consistently lead to better performance. This indicates that training on larger batches enables the LLM to learn stronger ranking abilities that generalize better to unseen data, which aligns with the theoretical analysis in~\cite{10.1145/1553374.1553449}. However, we also observe that larger batches make training more difficult to converge. This is intuitive, as ranking becomes more challenging with longer lists, increasing the difficulty of optimization. Therefore, how to effectively scale up listwise ranking through large-batch training or parallel optimization may be a promising direction for future research.

\subsection{Effect of Base LLM}
To evaluate the impact of the base model in PaperEval, we conduct a comprehensive comparison using various LLMs on the NAID dataset. As shown in Table~\ref{table:LLM}, we make the following observations: 1) LLaMA* and Qwen consistently deliver the strongest performance across all evaluation metrics. 2) Interestingly, LLaMA*, which undergoes additional training on language-related tasks, significantly outperforms the base LLaMA. This suggests that such targeted training enhances the model’s capacity to comprehend the nuanced semantics inherent in scientific literature. 3) Although Phi, the smallest model in our comparison, struggles to retrieve top-ranked papers, it achieves competitive accuracy in overall ranking consistency.

\begin{figure}
\centering
\includegraphics[scale=0.55]{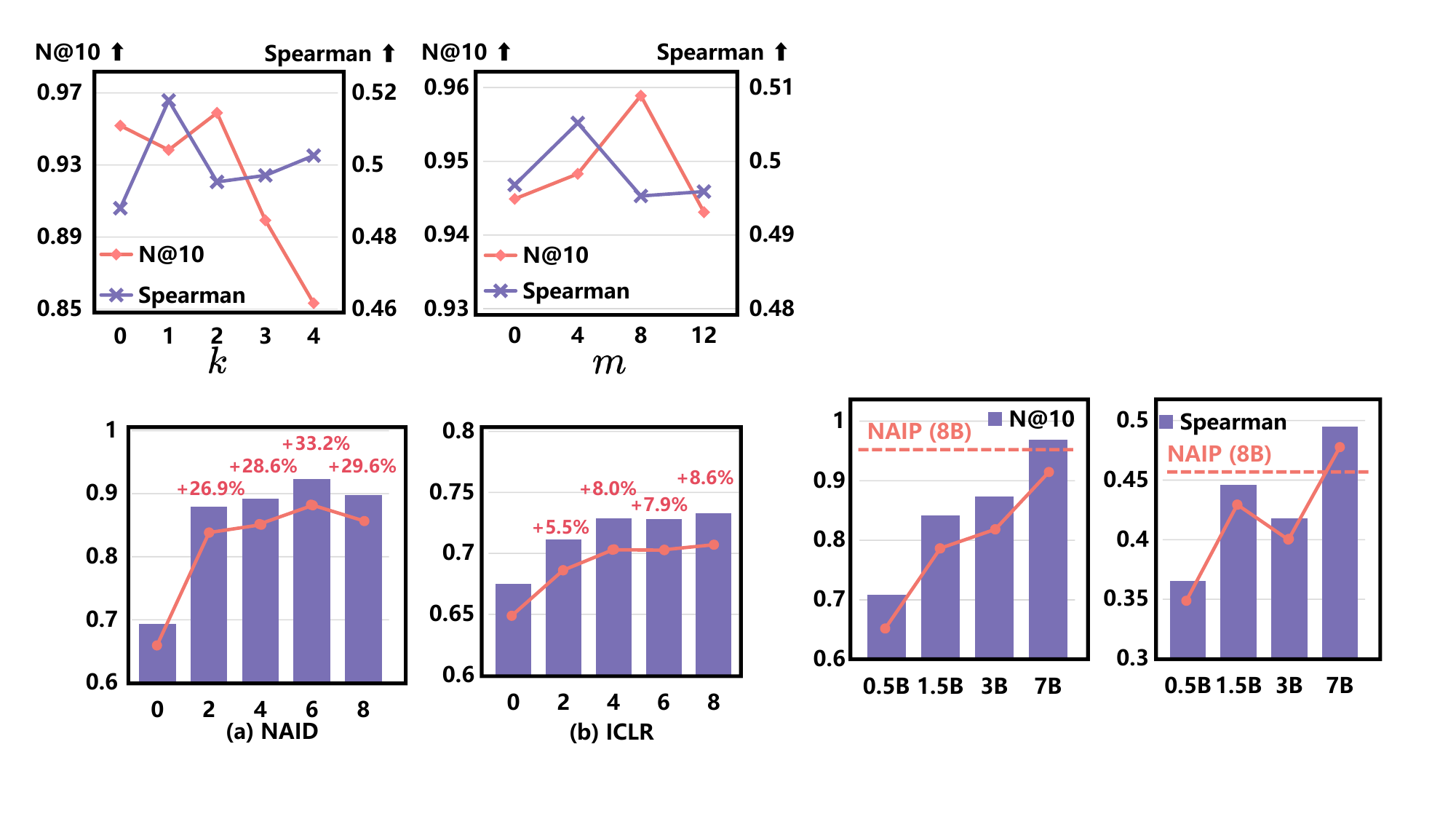}
\caption{Performance comparison across different sizes of Qwens used as base models. The dotted line is the performance of NAIP~\cite{zhao2025words}, based on 8B Llama3-Smaug~\cite{pal2024smaug}.}
\label{fig:scaling}
\end{figure}

We further investigate the impact of base LLM size on performance. We choose Qwen2.5-Instruct~\cite{qwen2025qwen25technicalreport} as our base model family due to its strong performance even at smaller scales. As shown in Figure~\ref{fig:scaling}, performance consistently improves as the model size increases. This suggests that our method scales effectively with larger LLMs and that the overall performance is closely tied to the capabilities of the underlying base model.

\subsection{Effect of Temperature Decay}
As illustrated in Equation (3), we employ a linearly decreasing temperature schedule. A lower temperature makes the distribution sharper, leading to a larger loss as shown in Equation (5). Formally, the gradient with respect to the scores is scaled by a factor of $\frac{1}{\tau}$. This scaling effect intensifies the model's learning signal for refining score rankings as the temperature decreases, thereby enforcing stricter supervision on the relative orderings.

Furthermore, we explored additional temperature scheduling strategies and compared them with the scenario where no temperature decay is applied. Specifically, we implemented three alternative schedules as follows:
\subsubsection{Exponential Decay.}
\begin{equation}
\tau^{(j)} = \tau_{\text{max}} \left( \frac{\tau_{\text{min}}}{\tau_{\text{max}}} \right)^{\frac{j}{m}}
\end{equation}

\subsubsection{Cosine Decay.}
\begin{equation}
\tau^{(j)} = \tau_{\text{min}} + (\tau_{\text{max}} - \tau_{\text{min}}) \cos\left( \frac{j}{m} \cdot \frac{\pi}{2} \right)
\end{equation}

\subsubsection{W/o Decay.}
For comparison, we also delete the temperature decay design. Instead, all temperatures are set to the same $\tau_{\text{max}}$.

\begin{figure}
\centering
\includegraphics[scale=0.49]{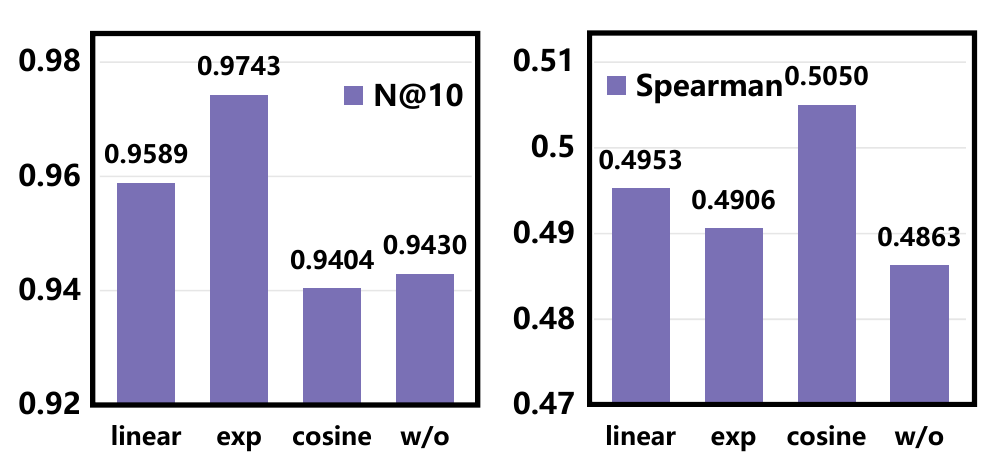}
\caption{Performance comparison across different temperature schedule designs.}
\label{fig:schedule}
\end{figure}

From the results presented in Figure~\ref{fig:schedule}, we can make the following observations: 1) The exponential decay schedule causes the temperature to decrease with an increasingly flattened rate. This manifests as the model gaining an advantage in top retrieval performance, yet showing limited improvement in overall ranking quality. 2) The cosine decay schedule leads to an increasingly accelerated drop in temperature. While this results in a significant enhancement in overall ranking quality, its performance in top-K ranking quality is even inferior to that without any decay. 3) The linear decay schedule reduces the temperature at a constant rate, which is reflected in notable improvements in both top-K ranking quality and overall ranking consistency.

\subsection{Computational Cost}
Our setup matches NAIP, and both retrieval and latent reasoning add minimal overhead while yielding significant gains. With batch size 64, inference takes 12.3s/batch (30 GB VRAM). In PaperRec, all new papers are evaluated within 2 minutes daily, confirming the method’s efficiency.

\subsection{Prompts}
In this section, we present all the prompts used in our paper.

\begin{tcolorbox}[colback=gray!5!white, colframe=gray!70!black, title=Topic extraction]
Given the title and abstract below, determine the specific research field by focusing on the main application area and the key technology. You MUST  respond with the keyword ONLY in this format: xxx.\\
Title: \{paper title\}\\
Abstract: \{paper abstract\}
\end{tcolorbox}

\begin{tcolorbox}[colback=gray!5!white, colframe=gray!70!black, title=PaperEval for NAID]
Given a certain paper. \\ 
Title: \{title\}\\ 
Abstract: \{abstract\}. \\ 
Predict its normalized academic impact (between 0 and 1)
\end{tcolorbox}

\begin{tcolorbox}[colback=gray!5!white, colframe=gray!70!black, title=PaperEval for ICLR]
You are a professional paper reviewer, given a certain paper. \\
Title: \{title\}\\ 
Abstract: \{abstract\}. \\
Generate a score for the paper (between 0 and 1)
\end{tcolorbox}

\end{document}